\documentclass[prd,superscriptaddress,twocolumn]{revtex4}

\usepackage{epsf,latexsym}
\usepackage{amssymb,amsmath}
\usepackage{times,graphics}
\usepackage[final]{pdfpages}

\usepackage{soul,xcolor}
\usepackage{epsfig}

\def\be{\begin{equation}}
\def\ee{\end{equation}}

\def\bes{\begin{align}}
\def\ees{\end{align}}

\def\bay{\begin{array}}
\def\ear{\end{array}}

\def\nn{\nonumber}

\def\etal{\textit{et al.}}

\def\1{{{\mathbbm 1}}}

\def\anab{{\mbox{\sf\boldmath{{$\nabla$}}}}}
\def\half{\mbox{$1\over2$}}

\def\pad{{\partial}}

\def\are{\mathsf{e}}

\def\ak{{\sf k}}
\def\au{\mathsf{u}}

\def\sg{\textsl{g}}

\def\bE{\mathbf{E}}

\def\bob{\mathbf{b}}
\def\boe{\mathbf{e}}
\def\bof{\mathbf{f}}

\def\bg{\mathbf{g}}
\def\bk{\mathbf{k}}

\def\hbb{{\hat{\bob}}}
\def\hbe{{\hat{\boe}}}
\def\hbf{{\hat{\bof}}}
\def\hbk{{\hat{\bk}}}
\def\hbl{{\hat{\mathbf{l}}}}

\def\hbw{\hat{\mathbf{w}}}
\def\hbz{\hat{\mathbf{z}}}

\def\hn{{\hat{n}}}

\def\vb{{\vec{b}}}

\def\vv{{\vec{v}}}
\def\vx{{\vec{x}}}

\def\vJ{{\vec{J}}}

\def\vom{{\vec{\omega}}}

\def\ak{{\sf k}}

\def\anab{{\mbox{\sf\boldmath{{$\nabla$}}}}}

\def\co{{\cal O}}

\newcommand{\Omegab}{\mbox{\boldmath$\Omega$}}
\newcommand{\omegab}{\mbox{\boldmath$\omega$}}

\begin{document}
\title{Post-Newtonian gravitational effects in quantum interferometry}

\author{Aharon Brodutch}
\affiliation{Institute for Quantum Computing, University of Waterloo, Waterloo, Ontario Canada N2L 3G1}
\affiliation{Department of Physics \& Astronomy, University of Waterloo, Waterloo, Ontario Canada N2L 3G1}
\author{Alexei Gilchrist}
\author{Thomas Guff}
\affiliation{Department of Physics \& Astronomy, Macquarie University, Sydney NSW 2109, Australia}
\author{Alexander R. H. Smith}
\affiliation{Department of Physics \& Astronomy, University of Waterloo, Waterloo, Ontario Canada N2L 3G1}
\affiliation{Department of Physics \& Astronomy, Macquarie University, Sydney NSW 2109, Australia}
\author{Daniel R. Terno}
\affiliation{Department of Physics \& Astronomy, Macquarie University, Sydney NSW 2109, Australia}
\email{daniel.terno@mq.edu.au}

\begin{abstract}
We investigate general properties of optical interferometry in stationary spacetimes and apply the obtained results focussing on quantum-optical experiments in near-Earth environments. We provide a rigorous expression for the {gravitationally induced} phase difference and adapt the parameterized post-Newtonian formalism for calculations of polarization rotation. We investigate two optical versions of the Colella-Overhauser-Werner experiment and show that the phase difference is independent of the post-Newtonian parameter $\gamma$, making it a possible candidate for an optical test of the Einstein equivalence principle. Polarization rotation provides an example of the quantum clock variable, and while related to the optical Lense-Thirring effects, shows a qualitatively different behaviour.
\end{abstract}

\maketitle
\setstcolor{red}

\section{Introduction}

Quantum theory and general relativity are often described as the two pillars of modern physics; this metaphor is apt in more than one way \cite{carlip01}. The two theories are built on different foundations --- probabilities that evolve in time cannot be easily reconciled with {a} deterministic unfolding of events in a dynamical spacetime. Their various aspects are verified with a spectacular precision, on  scales ranging from cosmic distances to   fractions of a millimetre {in the case of gravity} \cite{tur09}, and from $10^{-19}$m to 143 km for quantum mechanics  \cite{ma12}.

The early universe and late time black holes are two environments where both theories are expected to produce large effects. However, apart from an obvious lack of a  complete quantum theory of gravity  these regimes are hardly accessible for precision measurements.

Once a causal structure is fixed, relativity and quantum mechanics  {can} ``peacefully coexist" \cite{shimony78}.  In fact, explicit quantum effects and post-Newtonian gravity tend ``to share quarters by keeping clear of each other" \cite{ac14}. Two of the four classic tests of general relativity (light deflection and time delay) are based on classical interferometry \cite{sof,will93,will06}. Even if   quantum phenomenon, such as {the} M\"{os}sbauer effect  is used to test the weight of photons \cite{prebka,redshift}, the influence of gravity is modelled by  classical electromagnetic waves   on a curved background \cite{will93,will06}.

Moving atomic clocks \cite{wynads09} --- from the Hafele-Keating experiment \cite{hk72} to   GPS satellites are described as time-keeping devices moving along well-defined classical trajectories \cite{freqstan}. Moreover, when (non-relativisitc) quantum mechanics is used to describe  particle  motion, e.g, in neutron interferometry \cite{cow,cow2} or atomic fountains \cite{wynads09,wynwei},   Newtonian gravity is sufficient \cite{Gg,greenberger13}. This  {is} the case even if    matter-wave interferometry is used to bound   possible violations of   Einstein's equivalence principle \cite{chu11} in the gravitational Standard Model Extension \cite{sme}.

From   a theoretical point of view, there is a useful hierarchy of models  that describe  the interplay between quantum mechanics and gravity: beginning with non-relativistic quantum mechanics on a curved background (level 0), through quantum fields propagating on a curved background (level 1),  to semiclassical gravity (level 2), and  stochastic semiclassical gravity (level 3)  \cite{hu14},  and finally different effective field theory treatments of matter-gravity systems \cite{eft-g} and models based on the minimal length scale expected in the canonical quantum gravity and/or modified commutation relations \cite{sabine13}.

While it is generally accepted that the only modification at the {first} level  is due to obvious changes in the classical field modes on a given curved background, it {has} not been experimentally verified. The development of quantum technologies provides an opportunity to test explicit (and counterintuitive) quantum effects, like macroscopic superpositions and entanglement on a given gravitational background, and to investigate the interface between gravity and quantum mechanics. 

{The} quest for improved performance of   gravitational wave interferometers was one of the strong driving forces behind quantum metrology \cite{caves80}. Several quantum nondemolition \cite{braha} readout schemes for gravitational wave interferometers that are based on strongly non-classical states  {have been} investigated theoretically and experimentally. By  the end of this decade the second generation LIGO detectors may have been  upgraded enabling the injection of squeezed  light  \cite{adhikari14}. With this upgrade non-classical states on a non-trivial gravitational background    become experimentally accessible. In the field of atomic interferometry   phase calculations that  make use  {of} general relativity to model the effects of gravity \cite{aigr08} indicate that measuring non-Newtonian effects may soon become possible.  

Recent programs that aim to deploy quantum cryptography in space \cite{quest,sagas,qs12} provide a platform for single-photon interferometry sensitive to relativistic effects. A number of experiments  {have been} envisaged in this setting. The  QEYSSAT (Quantum EncrYption and Science SATellite) mission,  {led by the Institute for Quantum Computing in Waterloo, Canada, is being considered by the Canadian Space Agency.} Its main goal is a feasibility study of space-based quantum cryptography. In its basic configuration it consists of  {a} ground station, where  {a} photon source is located, and a microsatellite in a noon-midnight low-Earth orbit. Despite its simplified nature,    {the} QEYSSAT  {mission} will also allow the testing of long distance entanglement (over 1000 km).  {In addition,} a variety of quantum optics experiments  {have been proposed} that will test different aspects of  {the relationship between} quantum theory and relativity  \cite{qs12}; they range from  experiments achievable with current technology, to clearly fantastic. 

One of the proposed  experiments --- the optical version of the {Colella-Overhauser-Werner (COW)} experiment \cite{cow}  --- is considered in this article.  { In this experiment, a large Mach-Zender interferometer (MZI) is constructed and a gravity-induced phase, resulting from the arms of interferometer passing through different gravitational potentials, is measured}. While  {a} back-of-the-envelope estimation of the resulting phase shift can be made on the basis of the mass-energy equivalence and coupling of the resulting fictitious particle to  Newtonian gravity \cite{qs12}, the general relativistic effects are more subtle.  {Superpositions of a photon traveling in different arms of the interferometer will experience different time delays; if this} difference is comparable with  {the} photon's coherence time, the visibility of the interference pattern will drop \cite{bz11}.

Gravity causes polarization to rotate. This frame-dragging effect is known as a gravitomagnetic/Faraday/Rytov-Skrotski rotation \cite{skrot,fayos,gravimag1,gravimag2,bt11}. It may be possible to exploit this rotation to observe frame-dragging effects, which are the last of the classical tests of general relativity yet to be performed with sufficient accuracy \cite{will93,will06}. The main difficulty with the usual experiments that aim to detect frame-dragging is the necessity to isolate a much larger geodetic effect due to the Earth's mass from the frame-dragging that is caused by its spin. However, in the case of gravity-induced polarization rotation the net rotation along a closed trajectory is insensitive to the geodetic effect. It might be possible to measure this polarization rotation  and optically test frame-dragging effects.

Hence our motivation is twofold.  {First, to}   investigate the interferometric phase in a stationary spacetime and adapt it to the near-Earth environment. While supporting and qualifying previous results, {the results presented here} can be used as a convenient starting point for more sophisticated post-Newtonian analysis and higher-level calculations that include the effects of quantum gravity. {Second, to}   study in detail polarization rotation, which is interesting and important in its own right. Moreover,  the recently proposed interference of quantum clocks  --- particles with evolving internal degrees of freedom --- provides an example of a quantum effect that cannot be explained without the general relativistic notion of time \cite{br12}.  We discuss to what extent polarization rotation can be considered as such a clock.

The rest of this article is organized as follows: in Sec.~\ref{Interferometry in a stationary spacetime} we briefly review the  COW experiment, geometric optics, and interferometry in a stationary spacetime, rigorously deriving the resulting phase shift; in Sec.~\ref{Interferometry in the PPN approximation} we examine the optical COW experiment in the near Earth environment giving concrete estimates for both the resulting phase shift and gravity induced polarization rotation; and in Sec.~\ref{discuss} we discuss our results and comment on future directions.

To make the exposition more transparent we adopt the following notation:  a four-vector $\ak$ has the components $k^\mu$, $\mu=0,1,2,3$; the spacetime metric has the signature $-+++$ and is denoted by $\sg_{\mu\nu}$; coordinates of an event are labeled by $x^\mu=(t,\vx)$, where $\vec{x}$ stands for three spacelike coordinates in any spacetime. A vector in  a three-dimensional Euclidian space is denoted as $\vec{a}$, and it is equipped with the usual Euclidian inner product $\vx \cdot\vec{y}=x^1y^1+x^2y^2+x^3y^3$. A spatial metric $\gamma_{ab}$, $a,b=1,2,3$ is derived from $\sg_{\mu\nu}$, and $\bk$ denotes a three-vector on this space. Accordingly, $\hbk$ is a unit three-dimensional vector and $\hn$ is a unit Euclidean vector. We define the coordinate distance as $r:=\sqrt{\vx\!\cdot\!\vx}$. Unless it is stated otherwise we use an Earth-centered inertial system.

\section{Interferometry in a stationary spacetime\label{Interferometry in a stationary spacetime}}

To motivate our work, we begin by introducing the optical COW experiment. Then we briefly review the main features of geometric optics on a stationary curved background, providing  equations that govern the propagation and polarization rotation.

\subsection{Optical COW experiment\label{Optical COW experiment}}

\begin{figure}[htbp]
\includegraphics[width=0.5\textwidth]{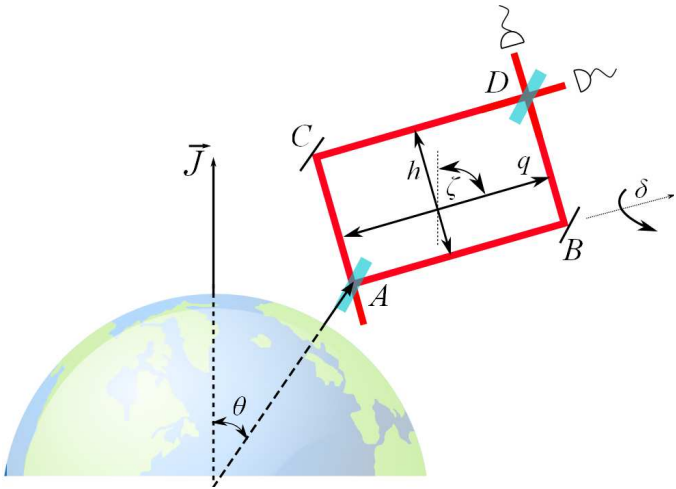}
\caption{Experimental setup: in the Earth-centred inertial frame (see Sec.~III) the $z$-axis is directed along the Earth's angular momentum $\vec{J}$. An interferometer is positioned in the $xz$-plane and oriented at an angle $\zeta$ with respect to the $z$-axis. The arms $AB$ and $CB$ have a length $q$ and are perpendicular to the direction of $g$ for $\zeta=\pi/2+\theta$. The arms $AC$ and $BD$ have length $h$. For the relationship between coordinates, directions, and physical distances see Sec. \ref{ppn:phase}. }
\label{setup}
\end{figure}

The COW experiment \cite{cow} was the first experimental demonstration of a gravitationally induced phase shift in a quantum system. Silicon crystals were used to split, reflect, and recombine a beam of neutrons, constructing a Mach-Zender interferometer.  The two paths taken by the neutrons in this MZI correspond to the arms $ABD$ and $ACD$ of Fig. \ref{setup}. By changing the inclination angle $\delta$ of the interferometer with respect to the horizontal plane, a range of phase shifts were obtained; with $\zeta=\pi/2+\theta$, the resulting phase difference is given by
\be
\Delta\psi=\frac{m^2g hq \sin\delta}{2\pi\hbar^2}=\frac{2\pi}{\lambda}\frac{g A\sin\delta}{v^2}, \label{cow1}
\ee
where $m$, $\lambda$, $v$ denote the neutron mass, de Broglie wavelength, and velocity respectively; $A$ denotes the area enclosed by the two arms of the interferometer and $g$ is the free fall acceleration at the surface of the Earth. Although there have been many improvements to the original experiment, a small discrepancy of 0.6$\%$-0.8$\%$ remains \cite{cow2}.

In the optical version of the COW experiment considered by QEYSSAT, an optical MZI is constructed on a scale sufficient for gravity to introduce an observable phase shift along one of the paths.  Such an experiment would involve sending a beam of coherent light through a beam splitter, with one of the sub-beams directly transmitted to a satellite and the other sub-beam sent through an optical delay at the Earth's surface before being transmitted. The two beams would be recombined at the satellite, completing the interferometer. The effect of the two sub-beams traversing paths with differing gravitational potentials, manifests itself as a phase shift in the resulting interference pattern, constituting a measurable effect of the gravitational redshift. The use of light rays instead of massive particles allows interferometry on the scale of $10^{5}$ m, which is not currently achievable with matter interferometry.

Making use of the equivalence of energy and gravitational mass $E=\hbar\omega\rightarrow m_{g}c^2$, and Eq.  \eqref{cow1}, one obtains a phase difference due to the gravitational redshift \cite{qs12}
\begin{equation}
\Delta \psi =\frac{2\pi}{\lambda}\frac{g hq}{c^{2}},\,  \label{eq:newtonphase}
\end{equation}
for a satellite located directly above a ground station, where $h$ is the height of the satellite, $q$ is the length of the optical delay, and $c$ is the speed of light. Using the estimated mission parameters of $\lambda=800$ nm, $h=400$ km, and $q=6$ km we get a significant phase shift of $\Delta \psi \sim \text{2 rad}$.

For the remainder of the article we consider a modified version of the QEYSSAT experiment pictured in Fig.~\ref{setup}, similar to the scheme considered  in \cite{br12}.

Semi-Newtonian arguments in the analysis of light propagation in a gravitational field qualitatively agree with the predictions of general relativity, but  may be off  by a factor of two \cite{will93, m1784,s1801}. In addition to the gravitational red shift, other effects, particularly the Doppler effect, are important. We give a careful derivation of the redshift effect in Sec.~\ref{ppn:phase}, and discuss other effects in Sec. \ref{discuss}.

\subsection{Geometric optics\label{Geometric optics}}

As appropriate for a level 1 model (quantum fields on a given classical background),  {light} propagation is described by the classical wave equation. Having in mind optical applications, we use the short-wave approximation \cite{bw}. Writing the complex vector potential as
\be
A^{\mu}=a^{\mu} e^{i\psi},
\ee
we assume \cite{bw,ll2,mtw} that the four-vector $a^\mu$ is a slowly-varying  amplitude that is independent of the wavelength $\lambda$, and the fast-varying phase (eikonal) $\psi$ scales inversely with $\lambda$.

The equations of geometric optics result from inserting this vector potential into the wave equation and imposing the Lorentz gauge.  {The} wave vector $k_\mu:=\pad_\mu\psi$, by definition is normal to hypersurfaces of constant phase $\psi$; {in addition} the wave vector is null, $k^{\mu}k_{\mu}=0$,  hence hypersurfaces of constant $\psi$ are null and their normals are also tangent vectors to the null geodesics $x^\mu(\sigma)$ they contain \cite{mtw, ep04}:
\be
\frac{dx^\mu}{d\sigma}=k^\mu, \qquad k^{\mu} \nabla_{\mu} k^{\nu} = 0. \label{mom1}
\ee
Here $\nabla_{\mu}$ is a covariant derivative compatible with the background metric $\sg$ and $\sigma$ is the affine parameter along a geodesic. Light rays are null geodesics that generate   surfaces of constant phase.

The eikonal equation, which is a restatement of the null condition, is given by
\be
\sg^{\mu\nu}\frac{\pad\psi}{\pad x^\mu}\frac{\pad\psi}{\pad x^\nu}=0, \label{HJequation}
\ee
{which} is the Hamilton-Jacobi equation for a free massless particle on a given background  {spacetime}. Hence when discussing classical electromagnetic wave propagation, it is sometimes convenient to refer to fictitious photons.

 The polarization vector is defined as $f^\mu:=a^\mu/\sqrt{a^\mu a_\mu^*}$, is transversal to the  {null} geodesic  {generated by $k^{\mu}$ and is parallel-propagated along it}
\be
f^\mu k_\mu=0,\qquad k^\mu\nabla_\mu f^\nu=0. \label{pol1}
\ee

 {Thus} we treat photons as massless point particles that move on the rays prescribed by geometric optics; the subtleties of   photon localization \cite{phtoloc} are not relevant for the discussion.  {In the geometrical approximation, } polarization can be described either in terms of a two-dimensional Hilbert space or   complex three- or four-dimensional vectors, which are orthogonal to the momentum \eqref{pol1} \cite{pt04,qs12}; here we adopt the latter.

\subsection{Phase\label{Phase}}

A key quantity in interferometry is the phase difference at the point of detection between {photons} that  take different paths in the interferometer \cite{bw}
\be
\Delta\psi(t,\vx):=\psi_{ABD}(t,\vx)-\psi_{ACD}(t,\vx),
\ee
where the subscripts refer to the path that the photon took.

Stationary spacetimes allow considerable simplification of the eikonal equation  due to the existence of a timelike Killing vector {field}, which allows one to define conserved energy (or frequency) of test particles \cite{ll2,mtw,ep04}. In particular, a Schwarzschild metric in isotropic coordinates \cite{mtw} takes the form
 \be
 ds^2=-V^2(r)c^2dt^2+W^2(r)d\vx\!\cdot\!d\vx,
 \ee
 where
 \be
 V=\left(1-\frac{r_{\sg}}{4r}\right)\bigg/\left(1+\frac{r_{\sg}}{4r}\right), \qquad W=\left(1+\frac{r_{\sg}}{4r}\right)^2,
 \ee
and the gravitational radius is $r_\sg:=2GM/c^2$. Its timelike Killing vector   is simply $\xi^{\mu} = \pad/\pad t$. Hence
 \be
k^{\mu} \xi_{\mu}=V^2k^0=-\omega_\infty/c=-V\omega_L/c=\mathrm{const},
 \ee
where $\omega_L$ is the frequency measured by a local stationary observer and $\omega_\infty$ is the frequency at infinity.

In stationary spacetimes, solutions of the Hamilton-Jacobi equation   \eqref{HJequation} for a free particle take the form
\be
\psi(t,\vx)=-\omega_\infty (t-t_0)+\omega_\infty S(\vx,\vx_0),
\ee
where $(t_0,\vx_0)$ is the starting point {of the photon's}  trajectory; the explicit form of the function $S(\vx,\vx_0)$ does not concern us here. At a given moment of the coordinate time the three-dimensional projections of the light rays are orthogonal to the surfaces of constant $S$,
\be
k_m\frac{dx^m}{dl}=0,
\ee
for any curve $\vx(l)$ that is contained in the surface $S(\vx,\vx_0)=\mathrm{const}$.

Different families of constant phase hypersurfaces correspond to different families of rays and different paths correspond to different initial momenta of  photons. We distinguish trajectories that pass through the intermediate points  $B$ or $C$ in Fig. \ref{setup}, as well as the related phases, by appropriate  subscripts. {All} time and phase differences are   calculated at the point $D$.

The constancy of phase along null geodesics implies
\be
\omega_\infty(t-t_0)=\omega_\infty S_{\Gamma}(\vx(t),\vx_0),
\ee
where the subscript $\Gamma$ indicates a path that corresponds to a particular set of initial conditions.  As a result, the phase difference between the two arms of the interferometer is
\begin{align}
\Delta\psi(t,\vx_D) &= \omega_\infty S_{ABC} (\vx_D,\vx_A)-\omega_\infty S_{ACD}(\vx_D,\vx_A) \nn \\
&=\omega_\infty(t_{ABD}-t_{ACD}) =\omega_\infty\Delta t, \label{phase}
\end{align}
where $\Delta t$ is the  {difference in} arrival  {coordinate} times of photons travelling along paths $ABD$ and $ACD$. The   {proper} time difference of a stationary observer at some point is related to the coordinate time difference at that point  {by the relation}
\be
 d\tau=\sqrt{\sg_{00}}dt,
\ee

In the case of the Schwarzschild spacetime, this yields a difference in the proper time given by $\Delta \tau=V \Delta t$.The {gravitationally-induced} phase difference is then
\be
\Delta\psi=\omega_L\Delta\tau.
\ee
This result is correct regardless of the validity of the post-Newtonian approximation.

\subsection{Polarization\label{Polarization}}

Stationary spacetimes allow a convenient three-dimensional representation of the evolution of the polarization vector \cite{fayos,bdt11}. Static observers follow the congruence of timelike Killing vectors that  define a projection from the  spacetime manifold $\cal{M}$ onto a three-dimensional space $\Sigma_3$, $\pi:\mathcal{M}\rightarrow \Sigma_3$. In practice, this projection is performed by dropping the timelike coordinate of an event, and  vectors are projected via a push-forward map: $\pi_*\ak=\bk$.  For a static observer, the three spacelike basis vectors of a local orthonormal tetrad are projected into an orthonormal triad: $\pi_*\are_{m}= \hbe_{m}$, with the property $\hbe_m \cdot \hbe_n=\delta_{mn}$.

The metric $\sg_{\mu \nu}$ on $\mathcal{M}$ can be written in terms of a three-dimensional scalar $h$, a vector $\bg$ with components $\sg_m$, and a three-dimensional metric $\gamma_{mn}$ on $\Sigma_3$ as
\be
ds^2=-h \left( dx^0 - \sg_m dx^m \right)^2+dl^2,
\ee
where $h:=-\sg_{00}$, $\sg_{m}:=-\sg_{0m}/\sg_{00}$, and the three-dimensional distance is  $dl^2:=\gamma_{mn}dx^mdx^n$, with
\be
\gamma_{mn}=\sg_{mn}-\frac{\sg_{0m}\sg_{0n}}{\sg_{00}}.
\ee
The three-dimensional $\gamma_{mn}$-compatible covariant derivative and the associated Christoffel symbols are denoted as  $D_m$ and $\lambda^m_{nl}$, respectively. Vector products and differential operators are defined as appropriate duals.

Using the relationships between the three- and four-dimensional covariant derivatives \cite{ll2}, the propagation equations \eqref{mom1} and \eqref{pol1} in a stationary spacetime, result in the following three-dimensional expressions  \cite{fayos,bt11,bdt11}:
\begin{align}
\frac{D\hbk}{d\sigma} &= \Omegab \times \hbk, \label{3dmrot}  \\
\frac{D\hbf}{d\sigma} &= \Omegab\times\hbf, \label{3dprot}
\end{align}
where $\sigma$ is an affine parameter along the curve with tangent vector $\mathbf{k} = d\mathbf{x}/d\sigma$. From Eqs. \eqref{3dmrot} and \eqref{3dprot} we see that both the propagation direction and polarization are rigidly rotated, with an angular velocity given by  \cite{bdt11}
\be
\Omegab=2\omegab-(\omegab\!\cdot\!\hbk)\hbk-\bE_g\times\bk, \label{Omegab}
\ee
with the vector $\left(h, \bg \right)^T$ playing the role of a vector potential for gravitoelectric and gravitomagnetic field \cite{inertia}:
\be
\bE_g=-\frac{\nabla h}{2 h}, \qquad  \omegab=-\half k_0 \nabla \times \mathbf{g}. \label{valom}
\ee

The polarization vector $\hbf$ is meaningful only with respect to a basis of two standard linear polarization vectors $\hbb_{x}$ and $\hbb_{y}$, or their superpositions $\tfrac{1}{\sqrt{2}}(\hbb_x\pm i\hbb_y)$ corresponding to right- and left-circular polarization vectors, for each momentum direction $\hbk$. Accordingly, the net polarization rotation along a photon's trajectory depends on the evolution of the standard polarization directions and is gauge-dependent \cite{bt11}.

In flat spacetime this basis is uniquely fixed by Wigner's little group construction \cite{wig,lpt1}. It is realized with the help of the conventionally defined standard rotations, which rotate a standard reference momentum to a particular direction. The reference momentum is chosen to be directed along an arbitrarily defined $z$-axis, with the reference frame being completed by the  $x$- and $y$-axes, which define the two standard linear polarizations $\hbb_{x}(\hbz)$ and $\hbb_{y}(\hbz)$. A direction  $\hbk$  is described by the spherical angles $(\theta,\phi)$. The standard rotation that brings the $z$-axis to $\hbk$ is defined as a rotation around the $y$-axis by $R_y(\theta)$, that is followed by a rotation $R_z(\phi)$ around the $z$-axis, so that $R(\hbk)=R_z(\phi)R_y(\theta)$. The standard polarizations vectors associated with the direction $\hbk$, are then defined as $\hbb_x(\hbk) := R(\hbk)\hat{\mathbf{x}}$ and  $\hbb_y (\hbk):= R(\hbk)\hat{\mathbf{y}}$.
%

However, on a general curved background the Wigner construction must be performed at every point, i.e., the standard polarization triad $(\hbb_x, \hbb_y,\hbk)$ is different at every location.

In the Schwarzschild spacetime, it is true that for any closed \textit{phase space} trajectory that a photon may travel,  the resulting phase difference $\Delta \chi$ between the the initial and final polarization vector, or equivalently the angle of rotation $\Delta \chi$ between the initial and final polarization basis, is zero \cite{skrot, gravimag1, gravimag2}; however, in general it is not true that this phase vanishes along each segment of the trajectory. This last statement is correct only for a particular gauge choice, which is constructed for a general (stationary) spacetime as follows: at each point in the spacetime we choose the standard reference momentum, or equivalently the $z$-axis of our polarization triad, to be aligned with the local free fall acceleration with respect to a static observer with four-velocity $\au$, that is $\hat{\mathbf{z}} := \hat{\mathbf{w}}$ where $\mathbf{w} = -\pi_*(\anab_\au\au)$; then for a photon with wave vector $\mathbf{k}$, we choose the linear polarization vector $\hat{\mathbf{b}}_y$ to be pointed in the direction $\hat{\mathbf{z}} \times \hat{\mathbf{k}}$, and finally we choose $\hat{\mathbf{b}}_x$ such that it completes the orthonormal triad $( \hat{\mathbf{b}}_x, \hat{\mathbf{b}}_y, \hat{\mathbf{k}})$. This construction is known as the Newton gauge \cite{bt11,bdt11}.

In the Schwarzschild spacetime the local free fall direction $\mathbf{w}$ is given by the gravitoelectric field $\mathbf{E}_g$, and the gravitomagnetic field vanishes. Thus, from Eq. \eqref{Omegab} we see the angular velocity, at which the polarization $\hbf$  and the propagation direction $\hbk$ of a photon rotate, is given by $\boldsymbol{\Omega} = - \mathbf{E}_g \times \mathbf{k}$, and points along the $\hbb_y$ direction. Such a rotation does not introduce a phase $\Delta\chi$ \cite{bt11}, which can be seen as follows. In the Schwarzschild spacetime, if a photon propagates from point $A$ to point $B$ in the plane defined by $\hbk$ and $\hat{\mathbf{b}}_x$, both its momentum $\hbk$ and linear polarization vector $\hbb_x$ are rotated around the direction $\hbb_y$, which is perpendicular to the plane of motion. This rotation, is exactly the same as the rotation relating the polarization triad at point $A$ to the polarization triad at point $B$, and consequently no phase in the polarization vector is acquired.

In addition to being defined by local operations, the Newton gauge has two useful properties: first, while it does not rely on a weak field approximation to define the reference directions, it considerably simplifies the post-Newtonian  calculations, as we show in Sec. \ref{Polarization rotation}; and second, if the trajectory is closed or self-intersecting, the reference direction $\hbw$ is the same at the points of the intersection.

Suppose we consider two different beams of light starting at point $A$, diverging,  and meeting again at point $B$, resulting in the propagation directions $\hbk_1$ and $\hbk_2$ at $B$ --- such as the typical situation in interferometry; even with a common reference direction $\hbw$ for the two beams, there are several useful ways to quantify polarization rotation. One simple way, which is employed in Sec. \ref{Polarization rotation}, is to compare the inner product of the polarization vectors of each beam at $B$, which directly influences the resulting interference pattern \cite{bw}, with the inner product of the initial polarization vectors at $A$. 

\section{Interferometry in the PPN approximation\label{Interferometry in the PPN approximation}}

The parameterized post-Newtonian (PPN) approximation \cite{sof,will93,mtw,inertia,wei} is  a systematic  method  for studying a system of slowly moving bodies bound together by  weak gravitational forces. This method admits a broad class of metric theories of gravity,   including general relativity as a special case, and is particularly well suited for Solar system experiments. In this regime $GM/r \sim v^2$, where $v$ is the velocity of a massive test particle or of some component of a gravitating body, is small. The PPN formalism constitutes a method for obtaining corrections to the Newtonian motions of a system, resulting from the metric theory in question, to higher orders in $GM/r$. The order of expansion is conveniently labelled by the parameter $\epsilon$, that is taken to unity at the end of the calculation \cite{mtw}. The scale is set by the upper bound on the gravitational potential, which is defined to be of the order $\epsilon^2$, e.g. at the centre of the Sun the gravitational potential is $\sim 10^{-5}$.

While performing calculations in the PPN formalism we will treat all vectors and vector products as Euclidian. Note that unit vectors such as $\hat{\mathbf{f}}$ associated with the metric $\gamma_{ab}$, are not necessarily unit vectors in Euclidian space, $\vec{f} \cdot \vec{f} \neq 1$.

Electromagnetic radiation and massless particles are not affected by Newtonian gravity. The leading post-Newtonian contributions (corrections to trajectories, time differences, and phases) are of order $\epsilon^2$ in the PPN expansion \cite{sof,mtw,will93}; to take into account gravitomagnetic effects we need contribution up to $\epsilon^3$.

The PPN expansion of the metric of a single slowly-rotating quasi-rigid gravitating body, assuming that the underlying theory of gravity is conservative and does not have preferred frame effects \cite{will93,mtw}, up to terms of order $\epsilon^3$ is given by
\be
ds^2=-V^2(r)c^2dt^2+\vec{R}\!\cdot\!d\vx\, cdt+W^2(r)d\vx\!\cdot\!d\vx, \label{farfieldkerr}
\ee
where
\be
V(r)=1-\epsilon^2\frac{U}{c^2}, \qquad W(r)=1+\epsilon^2\gamma\frac{U}{c^2},
\ee
where the Newtonian gravitational potential $-U=-GMQ(r,\theta)/r\simeq-GM/r$  depends on the mass $M$ and higher multipoles \cite{sof,inertia} (we take $Q\simeq1$), and
\be
\vec{R}=-\epsilon^3 2(1+\gamma)\frac{G}{c^3}\frac{\vJ\times\vx}{r^3}.
\ee
where $\vec{J}$ is the angular momentum {of the rotating body}. To simplify our calculations we choose the $z$-axis to be along the direction of angular momentum.

The PPN parameter $\gamma$ determines how local straight lines are bent relative to far away asymptotic straight lines. In general relativity $\gamma=1$, which is confirmed experimentally with an uncertainty less than a fraction of a percent \cite{tur09,will06}.

To obtain the leading contributions to the phase {and polarization rotation}, the photon trajectories only need to be expanded up to $\epsilon^2$,
\be
\vx(t)=:\vx_{(0)}(t)+\epsilon^2\vx_{(2)}(t)=\vb+\hn c(t-t_0)+\epsilon^2\vx_{(2)}(t),
\ee
where $\vb$ is the initial coordinate and $\hn$ is the Newtonian (zeroth-order {in $\epsilon$}) propagation direction. The equations of motion become \cite{will93}
\begin{align}
&\frac{d^2 \vx_{(2)}^{\,\perp}}{dt^2}=(1+\gamma)\big (\vec\nabla U-\hn(\hn\!\cdot\!\vec\nabla U)\big), \\
&\frac{dx_{(2)}^{\,\shortparallel}}{dt} =-(1+\gamma)\frac{U}{c}, \label{pardel}
\end{align}
where the parallel and perpendicular components of the second-order term are defined as
\be
 x_{(2)}^{\,\shortparallel}:=\vx_{(2)}\!\cdot\!\hn, \qquad
 \vx_{(2)}^{\,\perp}:=\vx_{(2)}-\hn(\vx_{(2)}\!\cdot\!\hn).
\ee

The coordinate time that it takes light to travel from $\vb$ to $\vx$ in the gravitational field of a massive body is longer than it would in the Newtonian gravity,
\be
c\Delta t=|\vx-\vb|+(1+\gamma)\frac{GM}{c^2}\ln\frac{r+\vx\!\cdot\!\hn}{b+\vb\!\cdot\!\hn},
\ee
where the second order terms is the Shapiro time delay \cite{shapiro}.

\subsection{Phase\label{ppn:phase}}

The phase difference is obtained by substituting the above result into Eq.~\eqref{phase}. We introduce the coordinate difference
\be
\Delta l:=|\vx_C-\vx_A|+|\vx_D-\vx_C|-|\vx_B-\vx_A|-|\vx_D-\vx_B|.
\ee
As indicated on Fig. \ref{setup}, the interferometer lies in the $xz$-plane.  Take the coordinates of point $A$  {to be}
\be
\vx_A = (b\sin\theta,0,b\cos\theta)^T,
\ee
and the directions {along different arms of the interferometer, e.g. $AB$, ect., to be}
\begin{align}
\hn_{AB} &= \hn_{CD}=(\sin\zeta, 0 ,\cos\zeta)^T,  \\
 \qquad \hn_{AC}&=\hn_{BD}=(-\cos\zeta, 0, \sin\zeta)^T,
\end{align}
resulting in  {the coordinates of the two mires at $B$ and $C$, and the detector at point $D$ to be}
\begin{eqnarray}
&\vx_B=\vx_A+q \hn_{AB}, \qquad \vx_C=\vx_A+h\hn_{AC}, & \nn\\
& \vx_D=\vx_B+h\hn_{AC}. \label{4corners}&
\end{eqnarray}
 {In what follows} we assume $b\gg h,q$.

Assuming that the coordinates of Eq. \eqref{4corners} are exact up to second-order  {in $\epsilon$}, resulting in $\Delta l=0$, and taking the leading terms in the expansion over the inverse powers of $b$, we obtain
\begin{align}
\Delta_l t &:=\left.t_{ABD}-t_{ACD}\right|_{\Delta l=0} \nn \\
&\approx(1+\gamma)\frac{GM}{c^3}\frac{hq}{b^2}\big(\sin(\zeta-\theta)-\cos(\zeta-\theta)\big).
\end{align}
 Taking into account that the free fall acceleration near the surface is $\sg=GM/b^2$, we obtain in the leading order
\be
\Delta_l\psi=\omega (1+\gamma)\frac{g}{c^3}hq\big(\sin(\zeta-\theta)-\cos(\zeta-\theta)\big). \label{tlator}
\ee

A physical path difference is obtained by following the trajectories in a three-dimensional space. The length of a segment $AB$ is
\be
L_{AB}=\int_A^B\sqrt{\gamma_{mn}dx^mdx^n}=\int_{t_A}^{t_B}W\sqrt{\frac{d\vx}{dt}\!\cdot\!\frac{d\vx}{dt}}dt,
\ee
where the integration os performed along the trajectory $AB$. In the first PPN approximation it reduces to
\begin{align}
L_{AB} &=\int_{t_A}^{t_B}\left(1+\frac{U\big(r(t)\big)}{c^2}\right)\left(c+\frac{dx_{(2)}^{\,\shortparallel}}{dt} \right)dt \nn  \\
&= c \Delta t_{AB}-\frac{r_\sg}{2}\ln\frac{r_B+\vx_B\!\cdot\!\hn_{AB}}{r_A+\vx_A\!\cdot\!\hn_{AB}}.
\end{align}

To simplify the exposition we consider the case where the MZI is oriented so the $AB$ arm is perpendicular to the local free fall acceleration, i.e. $\zeta=\pi/2+\theta$. The travel times $t_{AC}$ and $t_{BD}$ differ by a term proportional to $h q^2/b^3$ up to second-order in $\epsilon$, which can be neglected relative to the difference in the time of flight along the horizontal segments. If $L_{AB}=L_{CD}$, then
\be
\Delta_{L}t=\frac{g hq}{c^2}, \qquad \Delta_L\psi=\omega \frac{g hq}{c^3}. \label{tfiber}
\ee
The factor of two between the results given in Eq.~\eqref{tlator} and \eqref{tfiber} corresponds to the difference in setting up the actual experiments, as we discuss in Sec.~IV

\subsection{Polarization rotation\label{Polarization rotation}}

A combination of general results and the PPN expansion considerably simplifies our analysis. The polarization vector may be expanded along each segment of the photon's trajectory as
\be
f^m(t)=f^m_{(0)}+\epsilon^2 f^m_{(2)}(t) + \epsilon^3 f^m_{(3)}(t) + \co(\epsilon^4),
\ee
where the $f^m_{(0)}$ term represents the flat spacetime contribution, which remains constant along the enitre trajectory. The evolution of the polarization vector along the trajectory is given by Eq.~\eqref{3dprot}. In order to evaluate the three-dimensional covariant derivative appearing in the left hand side of Eq.~\eqref{3dprot}, we must expand the three-dimensional Christoffel symbols $\lambda^m_{nl}$ as
\be
\lambda^m_{nl}=\epsilon^2\lambda^m_{nl(2)} + \co(\epsilon^4),
\ee
where we have made use of the fact that the third-order term, $\epsilon^3\gamma^m_{nl(3)}$, vanishes in the PPN coordinates \cite{will93}. We can now expand the left hand side of Eq.~\eqref{3dprot} up to third-order in $\epsilon$ as a function of coordinate time
\begin{align}
\frac{Df^m}{d\sigma} &=\frac{Df^m}{dt}\frac{dt}{d\sigma} \nn \\
&=\epsilon^2\left(\frac{df^m_{(2)}}{dt}+c \lambda^m_{ij(2)} n^i f^j_{(0)}\right)\frac{\omega_\infty}{c^2}+\epsilon^3\frac{df^m_{(3)}}{dt}\frac{\omega_\infty}{c^2}.
\end{align}
The right hand side of Eq.~\eqref{3dprot} may be expanded as
\be
(\vec{\Omega} \times \hat{f})^m=\epsilon^2\left(\vec{\Omega}_{(2)} \times \hat{f}_{(0)}\right)^m+\epsilon^3\left( \vec{\Omega}_{(3)} \times \hat{f}_{(0)} \right)^m + \co(\epsilon^4).
\ee

As a result of the above, the second and third order terms of the polarization vector evolve independently. In the Newton gauge, there is no rotation relative to the polarization basis at second-order in $\epsilon$. This is because at this order in the expansion  the PPN metric in Eq. \eqref{farfieldkerr} and the far field Schwarzschild metric are identical, and as was discussed in Sec. \ref{Polarization}, employing the Newton gauge in the Schwarzschild spacetime results in the vanishing net polarization on any segment of the trajectory. Hence the polarization at any point up to the order $\epsilon^2$ can be recovered from the metric and the photon  momentum at that point. 

From Eq.~\eqref{Omegab}, we may expand the third-order term as
\be
\left( \vec{\Omega} \times \vec{f} \right)^m_{(3)}=\left(\big(2\vom_{(3)}-(\vom_{(3)} \cdot \hat{k}_{(0)})\hat{k}_{(0)}\big) \times \hat{f}_{(0)}\right)^m,
\ee
where $\vec{\omega}_{(3)}$ is obtained by expanding Eq.~\eqref{valom}:
\be
\vom_{(3)} =\frac{\omega_\infty}{c}\frac{(1+\gamma)}{2}\frac{GJ}{c^3r^5}
\begin{pmatrix}
3xz \\
3yz\\
3z^2-r^2
\end{pmatrix},
\ee
where $x$, $y$, $z$ are functions of the coordinate time $t$ and describe the trajectory to zeroth-order in $\epsilon$. The resulting equation for the evolution of the third order contribution to the polarization vector is given by
\be
\frac{d\vec{f}_{(3)}}{dt}=\frac{(1+\gamma)}{2}\frac{GJ}{c^3r^5} \vec{\Xi}_{(3)}\times \hat{f}_{(0)}, \label{ABarm}
\ee
where   along the $AB$ and $CD$ arms of the interferometer is
\be
\vec{\Xi}_{(3)}=\begin{pmatrix}
\left(r^2-3 z^2\right)\sin\! 2 \zeta +3 x z \cos \!2 \zeta +9 x z \\
0 \\
\cos \!2 \zeta  \left(r^2-3 z^2\right)-3 \left(r^2+x z \sin \!2 \zeta -3 z^2\right)
\end{pmatrix}
\ee
and along the $AC$ and $BD$ arms
\be
\vec{\Xi}_{(3)}=-\begin{pmatrix}
\sin \!2 \zeta  \left(r^2-3 z^2\right)-3 x z \cos \!2 \zeta +9 x z \\
0 \\
\cos \!2 \zeta \left(r^2-3 z^2\right)-3 r^2+6 x z \sin \!\zeta  \cos \!\zeta +9 z^2
\end{pmatrix}.
\ee

To account for the effect the reflection of the photon's at mirrors $B$ and $C$ has on the polarization vector, we expand the inward pointing normal of mirror $B$ as
\be
\vec{l}_B =
\begin{pmatrix}
\sin\left(\zeta + \pi/4\right) \\
0 \\
\cos\left(\zeta + \pi/4 \right)
\end{pmatrix}
-\epsilon^2\frac{\gamma r_\sg}{\sqrt{2}r_B}
\begin{pmatrix}
1 \\
0\\
0
\end{pmatrix}+\co(\epsilon^4),
\ee
and mirror $C$ as
\be
\vec{l}_C =
-\begin{pmatrix}
\sin\left(\zeta + \pi/4\right) \\
0 \\
\cos\left(\zeta + \pi/4 \right)
\end{pmatrix}
+\epsilon^2\frac{\gamma r_\sg}{\sqrt{2}r_B}
\begin{pmatrix}
1 \\
0\\
0
\end{pmatrix}+\co(\epsilon^4).
\ee
We then make use of a simple model of reflection and its effect on the polarization vector: the polarization component that is directed along $\hbl$ is unchanged, and the component that is orthogonal to $\hat{\mathbf{l}}$ changes sign \cite{bw}. Explicitly, the reflected polarization vector $\hat{\mathbf{f}}_r$ is given in terms of the incident polarization vector $\hat{\mathbf{f}}_i$ by
\begin{align}
\hat{\mathbf{f}}_r = 2 \left( \hat{\mathbf{f}}_i \cdot \hat{\mathbf{l}} \right) \hat{\mathbf{l}} - \hat{\mathbf{f}}_i.
\end{align}
Since the corrections to the normals are of order $\epsilon^2$ and from the above equation we can see we will be projecting $\epsilon^3$ terms in the polarization vector along them, we only need their zeroth-order in $\epsilon$  expressions, i.e. the flat space normals.  Admittedly this is a simple model of reflection valid for an ideal mirror; properties of actual mirrors, described by their Mueller matrices that depend on the wavelenght, incidence angle and have only approximate symmetries \cite{polar}.

To highlight the effect of polarization rotation, the light that follows the path $ABD$ is chosen to be initially polarized along the direction $\hbb(\hbk_{AB})\simeq\hat{b}_y(\hat{n}_{AB}) = \hat{y}$, i.e., perpendicular to the plane of the interferometer, and the light that follows the path $ACD$ to be polarized in the plane of the interferometer along the direction $\hat{b}_x(\hat{n}_{AC})= -\cos \zeta \hat{x} + \sin\zeta \hat{z}$. Without the gravitomagnetic effects, the two beams $ABD$ and $ACD$, would remain with orthogonal polarizations and no interference would be observed.

For simplicity we choose $\zeta= \theta + \pi/2$, so that the $AC$ arm of the interferometer lies along the radial direction, and integrating Eqs. \eqref{ABarm} and  along the $ABD$ arm of the interferometer results in the polarization vector
\begin{align}
\vec{f}_{ABD} &=  \hat{f}_{ABD (0)} \nn \\ &+ \epsilon^3\frac{GJ \left(1+\gamma\right)}{8b^3c^3}
\begin{pmatrix}
  8 h \cos 2 \theta +q \eta\left(\theta \right)  \\
  0  \\
-8 h  \sin 2 \theta  + q \eta\left(-\theta\right)
 \end{pmatrix}  &+\co(\epsilon^4) ,
\end{align}
where $\eta \left(\theta\right) := 6\left(\cos 2 \theta + \sin 2 \theta\right) +5 \left( \cos 4 \theta + \sin 4 \theta \right) +5$; and along the $ACD$ arm
\begin{align}
\vec{f}_{ACD}& =  \hat{f}_{ACD (0)}\nn \\ & + \epsilon^3 \frac{GJ \left(1+\gamma\right)}{b^3c^3}
\begin{pmatrix}
  0 \\
2 q \cos \theta - h \sin \theta \\
0
 \end{pmatrix} 
 &+\co(\epsilon^4).
\end{align}
In the above equations we disregarded the $\epsilon^2$ order contributions to polarization, as they lie in the planes defined by the corresponding $\hat{k}$ and $\hat{f}$ and do not contribute to the quantities of interest.

As discussed at the end of Sec \ref{Polarization}, to observe the effect of polarization rotation we compare the inner product of the initial polarization vectors at point $A$, which vanish since the initial polarizations were chosen to be orthogonal, with the inner product of polarization vectors at $D$:
\begin{widetext}
\begin{align}
\hbf_{ABD}\cdot\hbf_{ACD}=W^2\vec{f}_{ABD} \cdot \vec{f}_{ACD} =& -\frac{(1+\gamma) G J}{8 b^3 c^3} \Big[\left(8 h +q \right) \sin \theta +\left(8 h-5 q\right) \cos \theta  + 5 q \left(\sin 3  \theta + \cos 3  \theta \right) \Big] +\co(\epsilon^4).
\end{align}
\end{widetext}
Choosing favourable values for $\theta$ and $\zeta$, and taking $q=100$~km and $h=10$~km, we find $\vec{f}_{ABD} \cdot \vec{f}_{ACD}$ is approximately $2\times 10^{-17}$, which corresponds to a rotation of the initial polarization vector by roughly 4 pico arcseconds.

\section{Implications}\label{discuss}

We now discuss   interesting possibilities that the effects considered above offer, as well as  experimental challenges that should be overcome in their measurement.

\subsection{Red shift and interferometric phase}
\subsubsection{Experimental issues}

The phase difference given in Eq.~\eqref{tfiber}, agrees with the results of \cite{qs12,br12}. This scenario corresponds to the optical COW experiment with the arms $AB$ and $CD$ realized as optical fibre delay lines. The condition $L_{AB}=L_{CD}$ can be, at least in principle, also realized with a free-space propagation by suitably adjusting the mirrors {at $B$ and $C$}. The factor of $(1+\gamma)\simeq2$ highlights the difference between the physical distance and the difference in coordinates. The standard time-delay analysis assumes that the sender and receiver are far from the gravitating body, while   some trajectories pass close to it \cite{sof,will93,mtw,shapiro}. In this case, some of the coordinate differences actually represent physical distances, and the resulting phase difference is given by expressions similar to Eq.~\eqref{tlator}.

The phase difference due to the gravitational redshift is not directly observable in either satellite or ground-based experiments; it needs to be separated from other   effects. Consider first a satellite-based experiment.  Our treatment is similar to the analysis that was performed for Gravity Probe A \cite{gpa,gpa-fin}.  For a photon that was emitted from the ground station {with frequency $\omega_G$ and detected at the satellite, the frequency {$\omega_S$} measured onboard is
\begin{align}
\frac{\omega_S-\omega_G}{\omega_G} =& -\frac{U_G-U_S}{c^2}-\frac{v_G^2-v_S^2}{c^2} \nn \\
&-\frac{(\vv_S-\vv_G)\cdot\hn_{GS}}{c}+\co(c^{-3}),
\end{align}
where $v_G$ and $v_S$ are velocities of the ground station and the satellite in the Earth-centered inertial frame, $-U_G$ and $-U_S$ are the Newtonian gravitational potentials at the ground station and satellite respectively, and $\hn_{GS}$ is the propagation direction at the ground station. Assuming the   satellite is in a low Earth orbit  and neglecting influence of the Earth's multipole moments gives
\be
\frac{\Delta\omega}{\omega}\simeq -\frac{gh}{c^2}-\frac{v_G^2-v_S^2}{c^2}-\frac{(\vv_S-\vv_G)\cdot\hn_{GS}}{c}, \label{deltaomega}
\ee
where $h$ is hight of the satellite above the ground, $g$ is the free fall acceleration, and $\omega=\omega_G\approx \omega_S$.  The first term on the right hand side {of Eq.~\eqref{deltaomega}} is the gravitational red-shift, the second term represents the time dilation due to the relative motion of the satellite and ground station, and the third term results from the linear Doppler effect, which is about $10^5$ times larger than the other terms. Since the experiment relies on accumulating phase along different paths, unlike Gravity Probe A, there is no obvious way to effectively cancel  the linear Doppler effect \cite{gpa}.

In addition, since $v\sim\sqrt{gR_\oplus}$, the time dilation dominates the redshift by a factor   of $(R_\oplus/h)\sim 20$. Making this ratio {of the order of} unity  was the reason for launching Gravity Probe A to an altitude of 10,000 km in a nearly vertical trajectory \cite{gpa-fin}.

To obtain an interference pattern one has to adjust the phase {difference} $\Delta \psi$ {between the arms of the interferometer}. In ground-based experiments \cite{cow, br12} this is achieved by rotating the interferometer in the vertical plane. The analysis of the proposed QEYSSAT trajectory \cite{tj:private} shows that quantum communication will be possible for a range of elevations and inclination angles naturally occurring along it, thus enabling it to collect data corresponding to different phase differences. However, to extract the signal from the noise, a sensitivity to phase changes smaller than $10^{-7}$ is required, as well as precise information of the satellite's position and velocity.

Since the area {enclosed by} arms of a ground-based MZI \cite{br12} is much smaller {than in a satellite experiment}, the expected phase difference (for $h\sim100$ m, $q\sim6$km) is $\Delta \psi\sim 10^{-4}$. However, the Doppler effect is negligible, and the main competing effects are caused by the Earth's rotation. The leading terms in the PPN metric of Eq.~\eqref{farfieldkerr} of an isolated, quasi-rigid Earth that rotates uniformly around its $z$-axis  become in the comoving coordinates
\begin{align}
& V^2(r) =1- \frac{2U}{c^2}+\frac{\Omega_\oplus^2(x^2+y^2)}{c^2}, \\
& \vec{R} =- 2(1+\gamma)\frac{G}{c^3}\frac{\vJ\times\vx}{r^3}-2\frac{\vx\times\vec{\Omega}_\oplus}{c}, \\
& W^2(r) =1+ 2\gamma\frac{U}{c^2},
\end{align}
where $\Omega_\oplus$ is the angular velocity of Earth's rotation.

Orienting plane of the MZI along a meridian eliminates the Sagnac effect \cite{sag} (we discuss it in Sec.~IVB), while near the surface the redshift contribution from the Newtonian potential is larger than that of rotation by a factor of $300$.

\subsubsection{Applications}

As appropriate for the level 1 \cite{hu14} analysis, light propagation  {was} analyzed classically. A phase shift that is proportional to frequency leads to a reduction of the interferometric visibility, if the width of the wave-packet is not negligible. While this naturally can be explained within the theory of classical coherence \cite{bw},  {this analysis does not hold for } single-photon experiments that explore wave-particle duality \cite{br12}. Difference in flight times provide the which-way information (a particle property) and thus reduces the visibility (a wave property) \cite{es}.

Even without observing this decoherence, a MZI is a convenient platform for observing wave-particle duality. Indeed, if   a photon is first split by beam splitter $A$, travels inside an interferometer, and is finally recombined at a second beam splitter $D$ before detection; then if the second beam splitter is present, we observe interference fringes, indicating the photon behaved as a wave traveling in both arms of the MZI. If {on the other hand,} the beam splitter at $D$  is absent, we randomly register a click in only one of the two detectors, concluding that the photon travelled along a single arm, indicating particle properties.

In Wheeler's delayed-choice experiment \cite{wdc} one randomly chooses whether or not to insert the second beam splitter when the photon is already inside the interferometer and before it reaches $D$. The rationale behind the delayed choice is to avoid a possible causal link between the experimental setup and the photon's behaviour: the photon should not ``know" beforehand if it has to behave like a particle or like a wave. The choice of inserting or removing the beam splitter is either classically controlled by a random number generator \cite{jacq}, or by an axillary quantum system \cite{us}. The set-up we considered naturally lends itself to the largest scale delayed-choice experiment to date, as well as the phase shift resulting from an entirely different physical situation.

\medskip

Discussion so far assumed the exact validity of the Einstein equivalence principle (EEP) \cite{will93,will06,mtw}. Since  {the EEP} is a necessary element of any metric theory of gravity, the PPN formalism does not include a free parameter for the strength of its violation. From a phenomenological point of view, {we can introduce a parameter $\alpha$ that characterizes a possible violation, such that} {we can set}
\be
\sg_{00}=-1+(1+\alpha)U/c^2+\ldots,
\ee
 {which results in a modification of } the gravitational redshift   \cite{will06}
\be
\frac{\Delta\omega}{\omega}=(1+\alpha)\frac{\Delta U}{c^2}.
\ee
Experiments  {performed} in different systems, such as \cite{prebka,hk72,gpa-fin, chu11,red13}, resulted in bounds $|\alpha|\lesssim 10^{-6}-10^{-8}.$

The Standard Model Extension (SME) \cite{sme} is a framework to  {analize possible} violations of Lorentz symmetry. It takes the standard model of particle physics and adds a variety of tensorial quantities, combined with both fermionic and bosonic fields. The resulting terms in the action are not necessarily renormalizable and explicitly violate local Lorentz invariance. The latter is a key part of the EEP, {and thus} violations of the {EEP }are expected in such models. Numerous experiments are performed and re-interpreted \cite{smephen} to obtain  bounds on various Lorentz-violating terms. Experiments with atomic clocks that have led to the tightest bounds on $\alpha$, naturally focus on the fermionic sector and the associated Lorentz-violating terms \cite{chu11,red13}. In  fact, light is taken to propagate in the usual way \cite{chu11}.

The optical COW experiment offers a different system, where the quantum clock is realized by a single photon. Tests of light deflection and time delay are interpreted as giving a bound on $|\gamma-1|$ \cite{will93,will06}. However, the phase shift in the optical COW experiment Eq.~\eqref{tfiber}  is independent of $\gamma$, and hence the difference with general relativity can be directly interpreted in terms of the degree of violation {$\alpha$} of the EEP.

\subsection{Clocks and polarization}

Having additional internal degrees of freedom  {in a system which act like a clock, can} affect  the  {observed} interference   \cite{bz11}, if the evolution of this clock is affected by the  gravitational time dilation. While photons have no proper time,   polarization provides a similar clock. Unlike various clocks proposed for massive particles, polarization couples to gravity {similarly to the}  spatial degree of freedom. As with other clocks, the ability to control the internal degrees of freedom allows one to observe the interference that would be absent without gravity. In the case of polarization such an experiment can be sensitive only to the angular momentum contribution  {from} the gravitational field (Lense-Thirring effect), and different from the Sagnac interferometric experiments, where the gravitational contribution to the phase appear together \cite{sof}. However, this tuneability does not make estimating polarization rotation a viable option for optical measurement of the Lense-Thirring effect.

 {The} Sagnac effect \cite{sag}  {results in a} phase shift  {between} one counterpropagating wave with respect to another wave of the same mode in a rotating ring interferometer; the shift is proportional to the angular velocity of rotation $\vec{\Omega}$  {of the interferometer platform}, the area enclosed by the interferometer $\vec{A}$, and the frequency  {$c/\lambda$ of the wave:}
\be
\Delta\psi=\frac{8 \pi \vec\Omega\cdot\vec{A}}{\lambda \,c}.
\ee
Ring laser interferometers are very sensitive devices with a variety of uses, including a proposed measurement of post-Newtonian effects \cite{optgyro}. Approximating the Earth as a rigid sphere of a uniform density with a radius $R$, rotating at a constant angular velocity $\Omega_\oplus$, the   phase difference between the two waves of frequency $\omega$, including the leading PPN corrections, can be represented as \cite{sof}
\be
\Delta\psi=\frac{4\omega}{c^2} A\big(\Omega_\oplus\cos\zeta+\Omega_\mathrm{LT}+\Omega_\mathrm{G}\big)=:\Delta\psi_\mathrm{S}+\Delta\psi_{\mathrm{PN}},
\ee
where $\zeta$ is the angle between the normal to the interferometer and the $z$-axis. Both the Lense-Thirring and geodetic frequencies are proportional to $\Omega_\oplus r_\sg/R$, making the post-Newtonian contributions at least $10^{-9}$ times smaller than the classical Sagnac phase shift.

Polarization rotation behaves qualitatively different from the Sagnac effect: it does not scale with the area. As a result,
\be
\frac{\Delta\chi}{\Delta\psi_{\mathrm{PN}}}\sim\frac{l_+}{A}{c}{\omega}\sim\frac{\lambda}{l_-}\ll1,
\ee
where $l_-$ and $l_+$ are the short and the long arms of the MZI, and we assume that the area $A=l_+l_-$ enclosed by the interferometers is the same in both cases.

\section{Summary and perspectives}

We have examined several interesting effects present in both ground-based and near Earth space-based interferometry experiments. A rigours derivation of the gravitationally induced phase shift in the optical COW experiment was given, and concrete estimates of this phase along with the induced polarization rotation were derived. The sensitivity required to observe this effect is high, but experimentally accessible. Polarization rotation provides a realization of an internal clock for photons. However, the minuscule scale of the effect puts it beyond the current experimental reach. On the other hand, the optical COW experiment is a platform to test EEP with and for light. We plan to investigate the opportunities it provides in future work.

\medskip

\acknowledgments

We thank  Mile Gu, Bei-Lok Hu, Robert Mann, Tim Ralph, and Jayne Thompson for stimulating discussions and helpful comments,  and Roman Vander for advice on   ellipsometry. Fig.~1 uses the graphics by {\tt{barretr}} available at https://openclipart.org.

\medskip 
\appendix

\section{Comparison of polarizations}
The product of polarization vectors $\hbf_1\cdot\hbf_2$ gives one operationally meaningful way of quantifying polarization rotation along a closed trajectory. However, unless the directions $\hbk_1$ and $\hbk_2$ are collinear, some ambiguity remains. Beams can brought together in a variety of ways that are   described mathematically by rotation of one of the pairs $(\hbk_i,\hbf_i)$. Such rotations will in general lead  to different results for the polarization rotation.

The simplest  rotation of this type is performed in the plane defined by the vectors $\hbk_1$ and $\hbk_2$ that brings the two vectors together. Set $\hbz=\hbk_1\times\hbk_2/|\hbk_1\times\hbk_2|$, and define
\be
\hbb_y^i:=\hbz\times\hbk_i, \qquad \hbb_x^i:=\hbb_y^i\times\hbk_i=-\hbz.
\ee
As a result, the polarizations are written as
\be
\hbf_i=-f_i^x\hbz+f_i^y\hbb_y^i,
\ee
hence their product is
\be
\hbf_1\cdot\hbf_2=f_1^xf_2^x+f_1^yf_2^y\cos\varphi,
\ee
where $\varphi$ is the angle between the propagation directions.

After the  rotation $R_z^{-1}(\varphi)$ that brings $\hbb_2$ to $\hbb_1$ the standard polarizations also coincide. Hence the  product 
\be
\hbf_1\!\star\hbf_2:=f_1^xf_2^x+f_1^yf_2^y,
\ee
represents the polarization overlap after rotation,
\be
\hbf_1\!\star\hbf_2=\hbf_1\cdot R_z^{-1}(\varphi)\hbf_2,
\ee
giving an alternative comparator of polarizations.



\begin{thebibliography}{99}
\bibitem{carlip01} S. Carlip, Rep. Prog. Phys. \textbf{64}, 885 (2001).
\bibitem{tur09} S. G. Turyshev, Phys. Usp. \textbf{52}, 1 (2009).
\bibitem{ma12} X. S. Ma \etal, Nature \textbf{489}, 269 (2012).
\bibitem{shimony78} A. Shimony, Int. Phil. Quarterly \textbf{18}, 3 (1978); A. Peres and D. R. Terno, J. Mod. Opt. \textbf{49}, 1255 (2002).
\bibitem{ac14} G. Amelino-Camelia, Nature Phys. \textbf{10}, 254 (2014).
\bibitem{sof} M. H. Soffel, \textit{Relativity in astrometry, celestial mechanics and geodesy},  (Springer, Berlin, 1989).
\bibitem{will93} C. M. Will, \textit{Theory and experiment in gravitational physics}, (Cambridge University Press, 1993).
\bibitem{will06} C. M. Will, Living. Rev. Rel. \textbf{9}, 3 (2006).
\bibitem{prebka} R. V. Pound and G. A. Rebka, Jr., Phys. Rev. Lett. \textbf{4}, 337 (1960); R. V. Pound and J. L. Snyder, Phys. Rev. B788 (1965).
\bibitem{redshift} L.B. Okun', K. G. Selivanov, V. L. Telegdi, Phys. Usp \textbf{42}, 1045 (1999).
\bibitem{wynads09} R. Wynands, \textit{Atomic clocks}, in J.G. Muga, A. Ruschhaupt,  A. del Campo (eds.), \textit{Time in quantum mechanics, vol 2}, (Springer, Berlin, 2009), p.~363
    \bibitem{hk72} J. C. Hafele and R. E. Keating, Science \textbf{177}, 166 (1972); \textit{ibid.} \textbf{177}, 168 (1972).
\bibitem{freqstan}  F. Riehle, \textit{Frequency standards: basics and applications} (Wiley-VCH, Weinheim, 2004).
\bibitem{cow} R. Colella, A. Overhauser, and S. A. Werner, Phys. Rev. Lett. \textbf{34} 1472 (1975); J. L. Staudenmann, S. A. Werner, R. Colella, and A. W. Overhauser, Phys. Rev. A \textbf{21}, 1419 (1980).
\bibitem{cow2}  S. A. Werner, H. Kaiser, and R. Clothier, Physica B+C \textbf{151}, 22 (1988); H. Kaiser \etal, Physica B \textbf{385-386}, 1384 (2006).
\bibitem{wynwei} R. Wynands and S. Weyers, Metrologia \textbf{42},  S64 (2005).
\bibitem{Gg} A. Peters, K.Y. Chung, S. Chu, Nature \textbf{400}, 849 (1999); J.B. Fixler, G.T. Foster, J.M. McGuirk, M.A. Kasevich, Science \textbf{315}, 74 (2007).
\bibitem{greenberger13} W. P. Schleich, D. M. Greenberger, and E. M. Rasel, Phys. Rev. Lett. \textbf{110} 010401 (2013).
\bibitem{chu11} M. A. Hohensee, S. Chu, A. Peters, and H. M\"{u}ller, Phys. Rev. Lett. \textbf{106}, 151102 (2011).
\bibitem{sme} D. Colladay and V. A. Kosteleck\'{y}, Phys. Rev. D \textbf{55}, 6760
(1997); V. A. Kosteleck\'{y}, Phys. rev. D \textbf{69}, 105009 (2009);  V. A.  Kosteleck\'{y} and J. D. Tasson, Phys. Rev. D \textbf{83},
016013 (2011).
\bibitem{hu14} B. L. Hu, e-print arXiv/14026584 (2014).
\bibitem{eft-g} J. F. Donoghue, Phys. Rev. D \textbf{50}, 3874 (1994); C. P. Burgess, Liv. Rev. Rel. \textbf{7}, 5 (2004).
\bibitem{sabine13} S. Hossenfelder, Liv. Rev. Rel. \textbf{16}, 2 (2013).
\bibitem{caves80} C. M. Caves \etal,  Rev. Mod. Phys. 52, 341 (1980).
\bibitem{braha} V. B. Braginsky and F. Ya. Khalili, \textit{Quantum measurement}, (Cambridge University Press, 1992); V. B. Braginsky and F. Ya. Khalili, Rev. Mod. Phys. \textbf{68}, 1 (1996).
\bibitem{adhikari14} R. X. Adhikari, Rev. Mod. Phys. \textbf{86}, 1221 (2014).
\bibitem{aigr08}  S. Dimopoulos, P. W. Graham, J. M. Hogan, and M. A. Kasevich, Phys. Rev. D \textbf{78}, 042003 (2008).
\bibitem{lt-exp} Gravity Probe-B Science Results--�NASA Final Report
(2008); C.W. F. Everitt et al., Class. Quant.
Grav. \textbf{25}, 114002 (2008);
 I. Ciufolini and E. C. Pavlis, Nature  \textbf{431}, 958
(2004).


\bibitem{quest} R. Ursin \etal, Europhys. News \textbf{40}, 26 (2009).
\bibitem{sagas} P. Wolf \etal, Exp. Astron.  \textbf{23}, 651 (2009).
\bibitem{qs12} D. Rideout \etal, Class. Quant. Grav. \textbf{29}, 224011 (2012).
\bibitem{space} D. Bruschi \etal, Phys. Rev. D \textbf{90}, 045041 (2014).
\bibitem{bz11} M. Zych, F. Costa, I. Pikovski and \v{C}. Brukner, Nature Comm. \textbf{2}, 505 (2011).
\bibitem{skrot} G.V. Skrotskii, Sov. Phys. Dokl. \textbf{2}, 226 (1957).
\bibitem{fayos} F. Fayos and J. Llosa, {Gen. Rel. Grav.} \textbf{14}, 865 (1982).
\bibitem{gravimag1} H. Ishihara, M. Takahashi, and A. Tomimatsu, Phys. Rev.
D \textbf{38}, 472 (1988).
\bibitem{gravimag2} M. Nouri-Zonoz, Phys. Rev. D \textbf{60}, 024013 (1999); M.
Sereno, Phys. Rev. D \textbf{69}, 087501 (2004); S. M. Kopeikin and B. Mashhoon, Phys. Rev. D \textbf{65}, 064025 (2002).
\bibitem{bt11} A. Brodutch and D. R. Terno, Phys. Rev. D \textbf{84}, 121501(R) (2011).

\bibitem{br12} M. Zych \etal, Class. Quant. Grav. \textbf{29}, 224010 (2012).
\bibitem{m1784} J. Michell,  Philos. Trans. R. Soc. London \textbf{74}, 35 (1784).
\bibitem{s1801} J. G. von Soldner, in \textit{  Astronomisches Jahrbuch f\"{u}r das Jahr 1804}, p. 161  (F. S. E. Sp\"{a}then, Berlin, 1801); translation and comments in S. L. Jaki, Found. Phys. \textbf{8}, 927 (1978).



\bibitem{bw} M. Born and E. Wolf, \textit{Principles of optics}, 7th edition, (Cambridge University Press, 1999).
\bibitem{ll2} L. D. Landau and E. M. Lifshitz, \emph{The Classical theory of fields} (Butterworth-Heinemann, Amsterdam, 1980).
\bibitem{mtw} C. W. Misner, K. S. Thorn, J. A. Wheeler, \textit{Gravitation}, (Freeman, San Francisco, 1973).
\bibitem{phtoloc} O. Keller, Phys. Rep. \textbf{411}, 1 (2005); D. R. Terno, Phys. Rev A \textbf{89}, 042111 (2014).
\bibitem{ep04} E. Poisson, \textit{A relativisit's toolkit}, (Cambridge University Press, 2004).
 \bibitem{pt04} A. Peres and D. R. Terno, Rev. Mod. Phys. \textbf{76}, 93 (2004).
   \bibitem{bdt11} A. Brodutch, T. F. Demarie, and D. R. Terno, Phys. Rev. D \textbf{84}, 104043 (2011).

\bibitem{inertia} I. Ciufolini and J. A. Wheeler, \emph{Gravitation and inertia}, (Princeton University Press, 1995).
\bibitem{wig} E. Wigner, Ann. Math. \textbf{40} 149 (1939); S. Weinberg, \emph{The Quantum Theory of Fields} vol. 1 (Cambridge University Press, 1996).


\bibitem{lpt1}   N. H. Lindner, A. Peres, and D. R. Terno,  {J. Phys. A} \textbf{36}, L449 (2003).
\bibitem{wei} S. Weinberg, \emph{Gravitation and Cosmology: principles and applications of the general theory of relativity}, (John Wiley \& Sons, Inc., 1972).
    \bibitem{polar} J. Tinbergen, \textit{Astronomical polarimetry}, (Cambridge University Press, 1996); J. C. del Toro Iniesta, \textit{Introduction to spectropolarimetry}, (Cambridge University Press, 2004).
\bibitem{shapiro} I. I. Shapiro, Phys. Rev. Lett. \textbf{13}, 789 (1964).
\bibitem{gpa} D. Kleppner, R. F. C. Vessot, N. F. Ramsey, Astroph. Space Sci. \textbf{6}, 13 (1970); R. F. C. Vessot and M. W. Levine, \textit{Gravitational redshift space-probe experiment}, (NASA technical report NASA-CR-161409, 1979).
     \bibitem{gpa-fin} R. F. C. Vessot \etal, Phys. Rev. Lett. \textbf{45}, 2081 (1980).

\bibitem{tj:private} T. Jennewein, private communications.
\bibitem{es} B.-G. Englert, Phys. Rev. Lett. \textbf{77}, 2154 (1996).
\bibitem{wdc} J. A. Wheeler, in \textit{Quantum theory and measurement},
edited by J. A. Wheeler and W. H. Zurek (Princeton
University Press, Princeton, NJ, 1984), p. 182;
 A. J. Leggett, in \textit{Compendium of quantum physics}, edited
by D. Greenberger, K. Hentschel, and F. Weinert
(Springer, Berlin, 2009), p. 161.
\bibitem{jacq} V. Jacques et al., Science \textbf{315}, 966 (2007).
\bibitem{us} C\'{e}leri \etal, Found. Phys. \textbf{44},
576 (2014); R. Ionicioiu, T. Jennewein, R. B. Mann, and  D. R. Terno, Nature Comm. \textbf{5}, 4997 (2014).
\bibitem{red13} M. A. Hohensee \etal, Phys. Rev. Lett \textbf{111}, 050401 (2013).
\bibitem{smephen} D. Mattingly,  Liv. Rev. Rel. \textbf{8}, 5
(2005); V. A. Kosteleck\'{y}´ and N. Russell, Rev. Mod. Phys. \textbf{83}, 11
(2011).

\bibitem{sag} G. E. Stedman, Rep. Prog. Phys. \textbf{60} 615 (1997); G. B. Malykin, Phys. Usp. \textbf{43}, 1229 (2000).
\bibitem{optgyro} M. O. Scully, M. S. Zubairy, and M. P. Haugan, Phys. Rev. A \textbf{24}, 2009 (1981); G. E. Stedman, K. U. Schreiber and H. R. Bilger, Class. Quant. Grav. \textbf{20},  2527 (2003); F. Bosi \etal, Phys. Rev. D \textbf{84}, 122002 (2011).
\end{thebibliography}
\end{document}